\documentclass[a4paper, 11pt]{article}

\usepackage{amssymb}

\usepackage[nodots]{numcompress}

\usepackage{graphicx}
\usepackage{amssymb}
\usepackage{amsmath}

\usepackage{url}

\begin{document}

\title{Polynomial methods for fast Procedural Terrain Generation}

\author{
  Yann Thorimbert\\
  \texttt{yann.thorimbert@unige.ch}
  \and
  Bastien Chopard%\\
  %\texttt{bastien.chopard@unige.ch}
}

%\author[1]{Yann Thorimbert*}
%
%\author[1]{Bastien Chopard}
%
%
%\address[1]{\orgdiv{Centre of Computing Sciences}, \orgname{University of Geneva}, \orgaddress{ \country{Switzerland}}}
%
%
%\corres{*Yann Thorimbert, Battelle - batiment A, 7 route de Drize, CH-1227 Carouge, \email{yann.thorimbert@unige.ch}}

\maketitle

\begin{abstract}
A new method is presented, allowing for the generation of 3D terrain and texture from coherent noise. The method is significantly faster than prevailing fractal brownian motion approaches, while producing results of equivalent quality. The algorithm is derived through a systematic approach that generalizes to an arbitrary number of spatial dimensions and gradient smoothness. The results are compared, in terms of performance and quality, to fundamental and efficient gradient noise methods widely used in the domain of fast terrain generation: Perlin noise and OpenSimplex noise. Finally, to objectively quantify the degree of realism of the results, a fractal analysis of the generated landscapes is performed and compared to real terrain data.
\end{abstract}

%elsevier ou jcgt ?
%frequency analysis ? ==> cf article http://www.sciencedirect.com/science/article/pii/S0169555X02002222
%modification profonde - bien dire que apporte modification a methode de base

%\begin{keyword}
%%% keywords here, in the form: keyword \sep keyword
%Coherent noise \sep fractal terrain \sep random texture generation \sep fractal brownian motion \sep procedural content
%
%%% MSC codes here, in the form: \MSC code \sep code
%%% or \MSC[2008] code \sep code (2000 is the default)
%
%\end{keyword}

%%
%% Start line numbering here if you want
%%
%\linenumbers

%% main text
\section{Introduction}\label{sect:intro}

Procedural terrain generation (PTG) methods have grown in number in the last decades due to the increasing performances of computers; video games, movies and animation find obvious use of PTG, for terrain generation as well as for texture generation. However, a less evident use of PTG can be found in more practical domains, as for instance vehicle dynamics \cite{Dawkins2012} or military training \cite{Smelik2009,Smelik2010}, where accurate methods for emulating real terrain are of interest. More generally, one can also mention the use made of procedural coherent noise in the field of fluid animation \cite{Bridson2007,Narain2008}, which helps to improve performances of turbulence modeling. The main advantages of procedural noise compared to non-procedural noise generation is both an immensely decreased memory demand and an increased amount of content produced. The drawback usually occurs at two levels: accuracy, which reflects terrain realism, and performance, since the generation step may induce an additional processing effort compared to meshes which are simply loaded from files. What makes a PTG method more appropriate than another one is therefore dependent of the needed levels of performance and realism. For instance, PTG for video game usually targets performance rather than accuracy \cite{Olsen2004,Johnson} in order to run on home computers, as long as the produced terrain resemble -- at least superficially -- real ones. In the other hand, PTG used for vehicle simulation is highly demanding in terms of accuracy and less in terms of performance. Whatever the application, the capability of the method to describe a self-similar pattern is a crucial point; a central issue in PTG is the apparent fractal behaviour of many natural patterns that numerical models should mimic in order to produce realistic shapes \cite{Mandelbrot,Ebert,Peitgen}. It has to be noticed that, though not investigated in this study, some other approaches like stochastic subdivision \cite{Lewis1987} allow to produce realistic, non-fractal results.

Although the focus is put on terrain generation in this study, it is worth mentioning that the methods described in this paper directly apply to texture generation, as shown in Section \ref{sect:results}.

\subsection{Related works}
Most popular approaches for low-dimensional PTG include methods from the family of random midpoint displacement such as diamond-square algorithm \cite{Fournier82,Miller1986} and methods from the family of gradient noise, such as Perlin noise \cite{perlin85} and simplex noise \cite{Perlin2001}. As a result of the number of parameters influencing the generated terrain, noises from this family can be easily improved in terms of quality, as done for instance in \cite{ParberryTerrain}. While simplex noise is found to be clearly more efficient than Perlin noise for dimensions higher than 3, the difference in terms of performances is slighter for low dimensions (see Section \ref{sect:results:perfs}). Many recent efforts have been done to develop alternative methods that are reviewed here. The use of cellular automata \cite{Johnson} and tile-based procedural generation \cite{Grelsson2014} has shown to allow either fast, non-realistic terrain generation, either to be too slow for realistic-shaped height map generation. Evolutionary algorithms have been investigated to assist fractal terrain generation for video games, although not giving satisfactory results \cite{Raffe2012}. While tectonic-uplift \cite{Rankin2015}, hydrology-based \cite{uplift1,Genevaux2013} and example-based \cite{Zhou2007} approaches can reach a high level of realism, they are significantly slower than Perlin-like methods. In particular, the latter has serious limitations in terms of autonomous procedural generation, since it involves the synthesis of a template heightmap given by the user, whose features are transcribed in the generated map. Another promising approach based on terrain features examples can be found in \cite{guerin2016}, although no direct performance comparison with other methods can be found so far in the literature. Many other recent approaches, aiming at improving controllability and expressiveness of the noise, can be mentioned, as Gabor noise methods \cite{lagae2009,galerne2012}, random-phase noise \cite{Gilet2014} and wavelet noise \cite{Cook2005}, although not providing quantitative performance nor quality comparison between different existing models.

Midpoint displacement methods are usually more interesting than gradient noise methods in terms of pure performance, thanks to the fact that they do not need multiple passes corresponding to the different octaves. However, they suffer two serious drawbacks, namely their non-locality (the $h$ value at a given coordinate depend on its neighboring points) and the quality of the generated terrain, as pointed out by \cite{Miller1986}. Although most of the visual artifacts can be corrected by the more complex scheme presented in \cite{Lewis1987}, wich largely differs from the initial and simple idea of midpoint displacement, the fully local nature of algorithms such as Perlin noise allows to save a large amount of memory compared to non-local schemes and, moreover, is embarrassingly parallel. In addition, the nature of the midpoint displacement algorithms is such as it provides less control parameters than Perlin or simplex method. Finally, as for cellular automata, the non-locality of midpoint displacement causes additional difficulties for assembling different chunks of terrain, as commonly done in video games.

As a result of the specificities of the models discussed above, Perlin-like and simplex-like approaches clearly appear to be the most appropriate choice for fast generation of realistic terrains, thanks to the good compromise they provides between performance and quality. As pointed out by \cite{Gilet2014} and \cite{Cook2005}, Perlin noise method is still by far the most popular method due to combined effects of simplicity, performance, quality and historical inertia. To illustrate this fact, one can observe that most of the recent video games using massive procedural content generation make use of Perlin noise or variant; Minecraft (24 millions units sold between october 2011 and october 2016 \cite{minecraft}) or No Man's Sky (1.5 million units sold in the first three months \cite{noman}) to name but two (as a way of comparison, Tetris has been sold 30 million times since 1989 \cite{tetris}).

For all the reasons mentioned, this article focuses on comparisons of the presented model with Perlin and simplex noise in two dimensions, mapping two spatial quantities $(x,y)$ to a scalar value $h$. Note that OpenSimplex algorithm is used in this study to compare to the presented method, as it provides noise that is very similar to simplex noise and that it is not patented, unlike original simplex noise.

To conclude this review of existing methods, note that many algorithms as for instance cell noise \cite{Worley96} or erosion modeling \cite{Benes2002,Olsen2004,stava2008} are normally not used on their own but rather applied to results obtained by previously cited methods in order to increase accuracy. In addition, one can mention procedural generation of human structures like cities \cite{Parish2001,Dang2015}, which can be associated with terrain generation for human-impacted landscapes. Although these additional methods are not discussed here, it should be noted that the models described in this article are suitable for their use.

The aim of this study is to build a novel, general method based on boundary-constrained polynomials, which enclose Perlin noise, and to derive an optimized model for producing 2D heightmap using a minimum number of operations per pixel.

\section{Polynomial terrain generation model} \label{sect:model}

\subsection{Generalized boundary condition}
Consider a $D$-dimensional domain called cell, and restricted from 0 to 1 in each axis of the space: $x_a \in [0,1], \forall a \in [1,D]$. The position of a point in this cell is then characterized by a set of $D$ values $\mathbf{X}\equiv(x_1,x_2,..,x_D)^T$. The set of points $S$ for which all coordinates values are either 0 or 1 constitute the corner points of the cell. The cardinal number of $S$ in $D$ dimensions is equal to $2^D$.

Let $h$ be a function allowing to associate a height value $h(\mathbf{X})$ to each point of the domain. Here one wants to impose a height value $h(\mathbf{s}_i)$ to each corner point $\mathbf{s}_i\in S$. Similarly, values for spatial derivatives of $h$ can be imposed. Defining the partial mixed derivatives $h^{\mathbf{d}}$ as
\begin{equation}\label{eq:mixedderiv}
h^{\mathbf{d}}\equiv \frac{\partial^{d_1+d_2+..+d_D}}{ \partial^{d_1} x_1 ~\partial^{d_2} x_2~ ..~ \partial^{d_D} x_D}h,
\end{equation}
one writes $h^{\mathbf{d}}(\mathbf{X})=h^{\mathbf{d}}(\mathbf{s}_i)$ if $\mathbf{X}=\mathbf{s}_i$. Note that if the mixed derivative of order $n$ is continuous, then the mixed partial derivative is unaffected by the ordering of the derivatives. By defining as Equation (\ref{eq:mixedderiv}) the mixed partial derivative for a given $\mathbf{d}$ to be unique, the assumption is made of continuous mixed partial derivatives.

As a last requirement, one wants $h(\mathbf{X})$ to depend only on $h(\mathbf{s}_i)$ $h(\mathbf{s}_j)$ if $\mathbf{X}$ is located on the edge connecting the two corners $\mathbf{s}_i$ and $\mathbf{s}_j$, so that if this edge is shared with another cell, the interface between cells is smooth. In the sequels, this condition is referred to as the \emph{edge boundary condition}. Defining $\mathbf{\Delta}_{ij} \equiv \mathbf{s_j}-\mathbf{s_i}$, it reads:
\begin{equation}\label{eq:independance}
h\left(\mathbf{s_i}+k\cdot\mathbf{\Delta}_{ij}\right) = f(\mathbf{s_i},\mathbf{s_j}) ~\forall k\in[0,1] \Leftrightarrow \left\| \mathbf{\Delta}_{ij} \right\| = 1,
\end{equation}
where $f(\mathbf{s_i},\mathbf{s_j})$ must be a differentiable function of $\mathbf{s_i},\mathbf{s_j}$ only.

\subsection{Polynomial definition}
Many choices are possible at this point for the form of $h$; it is here described as a multivariate polynomial of degree $n$:
\begin{equation}\label{eq:polynomial}
h(\mathbf{X}) = \sum_{\mathbf{a}\in I} \left(c_\mathbf{a} \prod_{k=1}^D x_k^{a_k}\right),
\end{equation}
where $I$ is the set of all vectors on the form $(a_1,a_2..,a_D)$ such that $0\leq a_k \leq n ~\forall k$.

% It is important to stress that, for convenience, this definition of multivariate polynomial of degree $n$ is specific to this study. Here $n$ is the highest power to which a given variable is raised, and not the sum of the maximum exponent of each variable. This choice is motivated by the fact that on the edge between two adjacent corners, which is the boundary of interest constraining the problem, the contributions from all variables except one will vanish, letting us with a 1D polynomial of degree $n$. For this reason, if one wants the evaluation of the polynomial to differ on two edges along a given axis, then it needs contributions from mixed monomials whose total degree is higher than $n$.

A specific nomenclature may be defined for sake of clarity. Noting $m$ the order of the highest constrained derivative, one can refer to a given cell configuration by D$d$M$m$N$n$. For instance, D2M1N4 stands for a cell of dimension 2, constrained on value and on first derivative, and whose $h$ value is given by a polynomial of degree 4. Note that not all configurations make sense, as for instance D1M8N1, a polynomial for which the number of constraints is obviously higher than the number of coefficients.

%%version1
%Together with the value constraints and derivative constraints defined above, as well as the edge boundary condition (\ref{eq:independance}), expression (\ref{eq:polynomial}) leads to a system of linear equations which can be solved in order to obtain polynomial coefficients, as long as the number of constraints does not exceed the number of coefficients, whose value is the binomial coefficient $\binom{n+D}{D}$. The number of constraint can be viewed as the number of corners times the number of constraints per corner. Since the number of different derivatives of degree $m$ is the combinations with replacements of $D$ elements on $m$ length sequence, one can express the total number of constraints as:
%\begin{equation}\label{eq:nconstraints}
%2^D \cdot \sum_{i=0}^m \frac{(D+i-1)!}{i!(D-1)!} \leq \frac{(n+D)!}{D!n!}.
%%2^D \cdot \sum_{i=0}^m \binom{D-1+i}{i} \leq \binom{n+D}{D}.
%\end{equation}

%version2
Together with the value constraints and derivative constraints defined above, as well as the edge boundary condition Equation (\ref{eq:independance}), Equation (\ref{eq:polynomial}) leads to a system of linear equations which can be solved in order to determine polynomial coefficients, as long as the number of constraints does not exceed the number of coefficients, whose value is $(n+1)^D$. The number of constraints can be viewed as the number of corners times the number of constraints per corner. Since the number of different derivatives of degree $m$ is the combinations with replacements of $D$ elements on $m$ length sequence, one can express the total number of constraints as
\begin{equation}\label{eq:nconstraints}
2^D \cdot \sum_{i=0}^m \frac{(D+i-1)!}{i!(D-1)!} \leq (n+1)^D.
%2^D \cdot \sum_{i=0}^m \binom{D-1+i}{i} \leq \binom{n+D}{D}.
\end{equation}
A cell whose polynomial and constraints configuration obey this inequality is then guaranteed to obey the specified constraints on corners points.

%\subsection{Unidimensional polynomials}\label{case1d}
%As an example, we show here some results obtained for $D=1$. Despite its simplicity, the monodimensional case is of interest since the evaluation of the multivariate polynomial between two adjacent corners always yields a 1D polynomial. If one wants to constraint the first derivative, then it is clear that the degree of the polynomial must be 3 at least, as provided by (\ref{eq:nconstraints}). In the D1M1N3
%
%juste dire que trivial, enoncer resultats et relater a smoothstep, smootherstep et smoothestep

%%%%%%%%%%%%%%%%%%%%%%%%%%%%%% SECT:MODEL %%%%%%%%%%%%%%%%%%%%%%%%%%%%%%%%%%%%%%%%%%%%%%%%

\section{Special cases in two dimensions}
Three special cases of 2D polynomials are discussed in this section. First, the minimum polynomial that can generate 2D terrain with both height and gradient imposed is shown, the D2M1N3 polynomial. It is then briefly described how usual Perlin noise correspond to an order 5 polynomial, and it is finally show how to simplify D2M1N3 by assuming zero gradient on corners.

\subsection{D2M1N3 polynomial}
Restrict now to the case where $D=2$ and where the derivative are constrained up to order 1. In this case, the number of constraints provided by Equation (\ref{eq:nconstraints}) is equal to 12, and the smallest degree $n$ which allows the polynomial to respect the constraints is 3. In this configuration, Equation (\ref{eq:polynomial}) simplifies to
\begin{equation}
h(x,y) = \sum_{i,j} c_{ij}x^iy^j,
\end{equation}
where $i$ and $j$ are integers comprised between 0 and 3, usual axes names $x$ and $y$ now stands for $x_1$ and $x_2$, and $c_{ij}$ denotes polynomial coefficients $c_{i_1,i_2}$. Defining $f(x,y)$ and $g(x,y)$ as the $x$-component, respectively $y$-component of the gradient at position $(x,y)$, one obtains
\begin{equation}
f(x,y) = \sum_{i>0} i\cdot c_{ij}x^{i-1}y^j,
\end{equation}
and
\begin{equation}
g(x,y) = \sum_{j>0} j\cdot c_{ij}x^iy^{j-1}.
\end{equation}

One can define $h_{ij}$ to be the requested height on corner at coordinates $(i,j)$, so the height constraint reads $h(0,0) = h_{00}$, $h(1,0) = h_{10}$, and so on. Similarly, four conditions arise from the $x$-gradient conditions $f_{ij}$ and four from the $y$-gradient conditions $g_{ij}$, defining a system of twelve linear equations.

A possible solution for this system of equations is:
\begin{equation}\label{c00}
c_{00} = h_{00},
\end{equation}
\begin{equation}
c_{10} = f_{00},
\end{equation}
\begin{equation}
c_{22} = c_{33}=c_{32}=c_{23}=0,
\end{equation}
\begin{equation}
c_{20} = 3(h_{10}-h_{00})-2f_{00}-f_{10},
\end{equation}
\begin{equation}
c_{30} = f_{10}+f_{00}-2(h_{10}-h_{00}),
\end{equation}
\begin{equation}
c_{21} = 3(h_{11}-h_{01})-2f_{01}-f_{11}-c_{20},
\end{equation}
\begin{equation}
c_{31} = f_{11}+f_{01}-2(h_{11}-h_{01})-c_{30},
\end{equation}
\begin{equation}\label{c11}
c_{11} = h_{01} + h_{10} - h_{00} - h_{11} + f_{01}+g_{10}-g_{00}-f_{00}.
\end{equation}

Note that for symmetry reasons, non-diagonal terms $c_{ji}$ can all be deduced from $c_{ij}$ by inverting all indices and replacing $f$ by $g$. For instance, $c_{02} = 3(h_{01}-h_{00})-2g_{00}-g_{01}$. With these coefficients, it is easy to check that edge boundary condition is verified.

%%%%%%%%%%%%%%%%%%%%%%%%%%%%%% PERLIN %%%%%%%%%%%%%%%%%%%%%%%%%%%%%%%%%%
\subsection{Perlin's polynomial}
In Perlin's method, a grid of gradient values is generated and the height value of a subdomain is obtained by interpolating the height contribution of each corner's gradient. The smooth interpolation function $S$ can be of any form as long as $S(0) = 0$ and $S(1) = 1$. It is most common to use polynomial of order 3 (smoothstep) or 5 (smootherstep) \cite{Perlin2002}. Smoothstep reads $S_3(x) = 3x^2-2x^3$ and is the lowest order polynomial to provide zero-derivative at $x=0$ and $x=1$ along with the condition aforesaid. The height value of a point in the domain then reads:
\begin{equation}
h(x,y) = h_0(x,y) + S(y)\cdot \left(h_1(x,y)-h_0(x,y)\right),
\end{equation}
with
\begin{equation}
h_0(x,y) = v_{00}(x,y) + S(x)\cdot \left(v_{10}(x,y)-v_{00}(x,y)\right)
\end{equation}
and
\begin{equation}
h_1(x,y) = v_{01}(x,y) + S(x)\cdot \left(v_{11}(x,y)-v_{01}(x,y)\right),
\end{equation}
where $v_{ij}(x,y) = f_{ij}\cdot(x-i) + g_{ij}\cdot(y-j)$.

%A enlever: (preuve que 0 aux quatres coins)
%\begin{equation}
%h(0,0) = v_{00}(0,0) = f_{00} \cdot 0 + g_{00}\cdot 0
%\end{equation}
%\begin{equation}
%h(1,0) = v_{10}(1,0) + \left(v_{10}(1,0)-v_{00}(1,0)\right) = g_{10} \cdot 0
%\end{equation}
%\begin{equation}
%h(0,1) = v_{01}(0,1) = f_{01} \cdot 0
%\end{equation}
%\begin{equation}
%h(1,1) = v_{01}(1,1) + \left(v_{11}(1,1)-v_{01}(1,1)\right) = 0
%\end{equation}

With the used definition of multivariate polynomial degree, it appears that this scheme makes use of a polynomial of degree $2+s$, where $s$ is the degree of the chosen smoothstep polynomial. The consequent total order is at least 5, a value higher than for D2M1N3 polynomial presented above, which is of degree 3. However, due to the factorized form it offers, Perlin noise allows to gain computation steps compared to D2M1N3, resulting in better performances; Zero-gradient D2M1N3 presented below, in the other hand, will need less operations than Perlin's polynomial.% Performance and accuracy comparison of D2M1N3, zero-gradient D2M1N3 and Perlin's polynomial is provided in Section \ref{sect:results}.

\subsection{Zero-gradient D2M1N3 polynomial}
Forsaking the generality of D2M1N3 polynomial derived above, one can impose special gradient conditions in order to increase performances of the implementation of the polynomial. The condition reads $f_{ij} = g_{ij} = 0~\forall i,j$ and Equations (\ref{c00}-\ref{c11}) simplify, yielding the following expression for $h(x,y)$:
\begin{multline}\label{eq:zg}
h(x,y) = h_{00} + S_3(x)\Delta x  + S_3(y)\Delta y  +
				A\left[ S_3(x)\cdot y + S_3(y)\cdot x +xy\right],
\end{multline}
with $\Delta x = h_{10}-h_{00}$, $\Delta y = h_{01}-h_{00}$ and $A=h_{11}+h_{00}-h_{10}-h_{01}$, and as before $S_3$ is the third order smoothstep function. It is worth noting that the only cell-dependent terms in this expression are $h_{00}$, $\Delta x$, $\Delta y$ and $A$; this allows for important performance gain when using lookup tables for evaluating space-dependant terms (\emph{i.e} terms involving $x$ or $y$), since only cell-dependant terms have to be evaluated, as a result of their dependancy to boundary constraints, that are not known before terrain generation. The quality of the generated terrain is not affected in comparison with the generic version of the polynomial (see Section \ref{sect:results:visu}). Note that $S_3$ could be replaced by any higher order smoothstep function $S_i$, in a very similar way as in Perlin noise.

Zero-gradient D2M1N3 polynomial is the minimum configuration (\emph{i.e} the configuration that has the lowest number of coefficients) for smooth two-dimensional heightmaps. Indeed, $N=3$ cannot be reduced since no polynomial of degree lower than 3 can take arbitrary height and derivative values at boundaries. In addition, $M=1$ cannot be reduced by definition as one is seeking for smooth heightmaps, that are necessarily constrained on first derivative. Thus, the only way to reduce the number of coefficients is to impose special values at the constrained locations. Examining Eqs (\ref{c00}-\ref{c11}), it appears that setting $f_{ij}=g_{ij}=0$ allows to cancel two coefficients and reduce the expression of the other ones.

%constat d(dh/dx)/dy ~ 0. PLUTOT constat de discontinuite, puisque pas imposee par configuration

%comment l'expliquer ? quadratique, ne peut croiser qu'une fois x ==> au lieu de prendre valeurs entre -0.5 et 0.5, les prend entre (+-) (0 et 0.5) ==> ~ constant sur les droites
%...

\subsection{Other polynomials of interest}
The systematic approach described in Section \ref{sect:model} is general and can be applied for dimension 2 as well as any dimension $D$, although simplex-like algorithms have shown to be more efficient for high dimensions (see Section \ref{sect:intro}). Unidimensional equivalent to D2M1N3 is D1M1N3, which leads to smoothstep function $S_3(x) = 3x^2 - 2x^3$ if specific conditions $h(0) = 0$, $h(1) = 1$ and zero gradients at boundaries are required. A 3D equivalent to D2M1N3 configuration would be D3M1N3. In addition to the generality of dimension, constraints determine the quality of the generated terrain. If the accuracy of the terrain is a priority, one may consider to impose gradients constraints for higher orders. In particular, with a view to improve isotropy, one may consider to use $M=2$ and impose second order, mixed derivative gradients $\partial^2 h/\partial_x^2,~\partial^2 h/\partial_y^2,~\partial^2 h/(\partial_x \partial_y)$ in addition to first order gradients. This would lead to D2M2N4 and D3M2N4 models.

\section{Fractal noise}\label{sect:noise}
In order to produce convincing results, it is of crucial importance to observe and reproduce the apparent fractal nature of real terrains. This is achieved by adding together multiple layers of height maps commonly called \emph{octaves}, generated by the same method but which lower in amplitude $a_i$ as they grow in frequency $f_i$, with $i$ the number of the octave. More specifically, $a_i\propto a^{-i}$ and $f_i \propto f^{i}$. When $a \approx f \approx 2$, the amplitude of the deformation is always proportional to the scale at which it is applied; this is how self-similarity arise from the generated terrain. It is worth noticing that most authors refer to $1/f$ as the lacunarity and is then seen as the multipler of frequency between two successive octaves. Finally, $a$ is often called persistence. Note that in this study these parameters are chosen once and for all before data is produced and do not influe the performance of an implementation. For all the results presented in this article, base frequency is equal to 2 and lacunarity is equal to 0.5, unless otherwise specified.

A property of D2M1N3 makes it a particular model compared to Perlin polynomial and zero-gradient D2M1N3 presented below; while both of these have, by construction, either zero gradient or zero height on the corners of the domain, generic D2M1N3 can take any arbitrary value on these locations. For this reason, one may define an additional parameter $w$ which weights the value of gradient in order to tune the predominance of either height or gradient at the corners. For terrain generation, $w$ is typically a constant, since the scale (\emph{i.e} the octave) seems to have no influence on the slope of the added values, in first approximation. In the sequels $w = h_0/100$, where $h_0$ is the amplitude of polynomial at octave zero.

\subsection{Time cost}
The computational time $T$ for generating a height map using the method presented in this study is discussed here. %, which make use of constrained polynomials as described in Section \ref{sect:model}. % Note that we will keep constant contributions into mathematical expressions of the complexity in order to further interpret the performances differences between methods of the same complexity, in \ref{sect:results}.
Consider first the case of a single octave. For a domain of size $R$, the computational time of the evaluation of a polynomial over the domain is equal to the number of evaluation points times the cost of the evaluation of a single point. In $d$ dimensions it reads $T_{eval} = C_{eval}\cdot R^d$, with $C_{eval}$ the cost of a single point evaluation, whose value depends on the specific type of polynomial used. In addition, each polynomial coefficient needs to be computed once and for all before spatial evaluation. At a given octave level $i$, there are $2^{i\cdot d}$ different polynomials to initialize, leading to a creation cost equal to $T_{init} = C_{init}2^{i\cdot d}$, where $C_{init}$ is the time needed for deducing polynomial coefficients from boundary conditions, which again depends on the specific type of polynomial used. Note that $C_{eval}$ and $C_{init}$ are typically of the same order of magnitude.

\begin{figure}[htbp]
  \includegraphics[width=1.\textwidth]{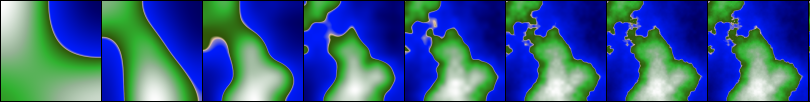}
	%\makebox[\textwidth][c]{\includegraphics[width=1.2\textwidth]{image}}%
  \caption{Example of map generated with 1 to 8 octaves, from left to right. Zero-gradient D2M1N3 model was used. As a result of the exponentially decaying height contribution, the global change of shape of the generated terrain, from the point of view of a given scale, is exponentially smaller.}
  \label{fig:octaves}
\end{figure}

The total cost $T=T_{eval}+T_{init}$, including the $N$ octaves, is simply the sum of the cost for each octave from 0 to $N$. One obtains $T \approx C_{eval} N R^d + C_{init}2^{N d}$. Let us now consider the two-dimensional case. The previous expression may be misleading, as the second term usually becomes negligible compared to the first one, despite the exponential behaviour of the latter, since dimension $d=2$, resolution $R\approx 10^3$ and $N\approx 7$ or smaller in most cases; in this configuration, non-exponential term is of the order of $10^7$ while the exponential one is of the order of $2^8 \approx 10^3$. Thus for usual, low number of octaves, the evaluation dominates the total cost of terrain generation, while polynomial initialization dominates the cost for high number of octaves. Performance test depicts this effect in Section \ref{sect:results:perfs}. This behaviour is related to the fact that the contribution of each octave exponentially decays. As a result, the global change of shape of the generated terrain, from the point of view of a given scale, is exponentially smaller. This can be observed in Figure \ref{fig:octaves} and is the reason why the exponential component of the time complexity can be neglected in most cases, since the typical resolution used is large enough to make it much smaller than the linear component up to approximately 7 octaves.

\section{Results} \label{sect:results}
%In all this section, we will use subscript $g$ for quantities related to generic D2M1N3 model, $z$ for zero-gradient D2M1N3 model, and $p$ for Perlin's model.

\subsection{Visual comparison with other methods} \label{sect:results:visu}

\begin{figure}[htbp]
  \centering
  \includegraphics[width=1.\textwidth]{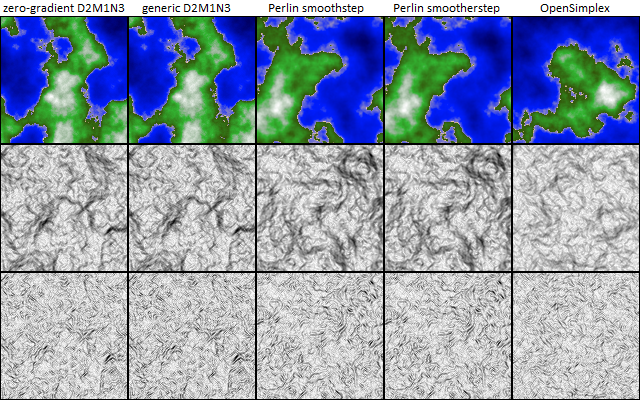}
  \caption{Example of terrain generated with generic D2M1N3, zero-gradient D2M1N3, third order smoothstep, fifth order smoothstep Perlin's model and OpenSimplex scheme on each column respectively. Height, height gradient norm and height second order gradient norm are represented on each line, for each corresponding column model. Specific color map was used to visualize terrain. Gray scale was used to visualize gradients, where white means zero and black means maximum value.}
  \label{fig:gradient}
\end{figure}

An example of terrain produced with different methods is shown in Figure \ref{fig:gradient}, along with its first and second order gradients. In addition to the three methods previously described, Perlin's model has also been used with a fifth order interpolation polynomial $S_5(x) = 6x^5 - 15x^4 + 10x^3$, as well as OpenSimplex method as a way of comparison. No quality difference can be visually noted between models; acceptable isotropy (\emph{i.e} not visible for human eye) at all orders can be observed, although it can be pointed out that simplex model provides values that are distributed slightly more evenly, as it can be seen from the first order gradient norm.

\begin{figure}[htbp]
	\centering
		\includegraphics[width=1.\textwidth]{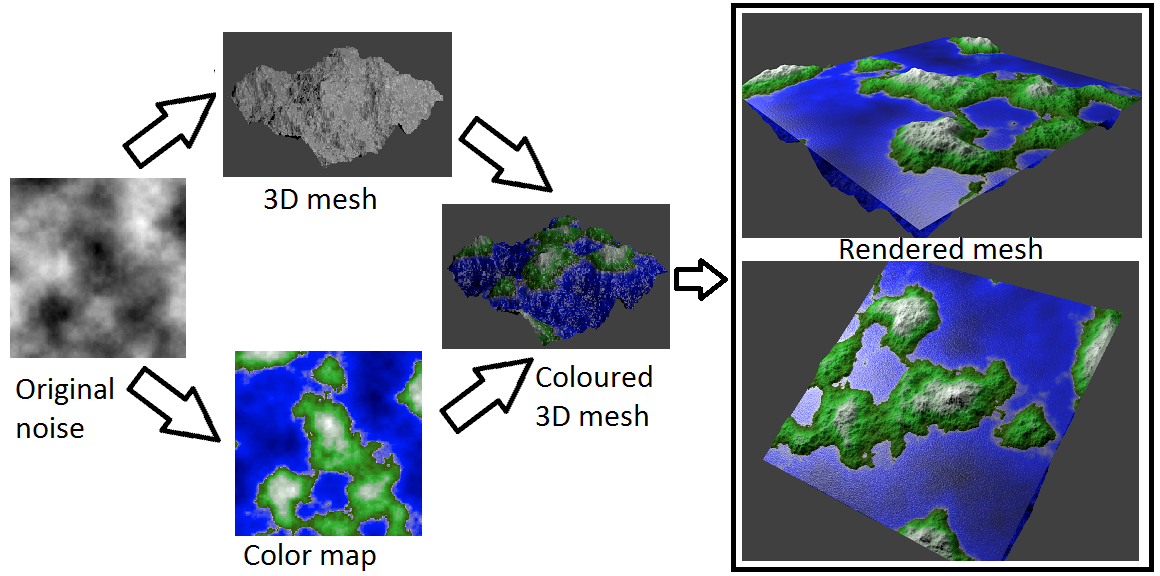}
	\caption{Process leading to 3D mesh, from raw noise to rendered mesh.}
	\label{fig:montage}
\end{figure}

\begin{figure}[htbp]
	\centering
		\includegraphics[width=1.\textwidth]{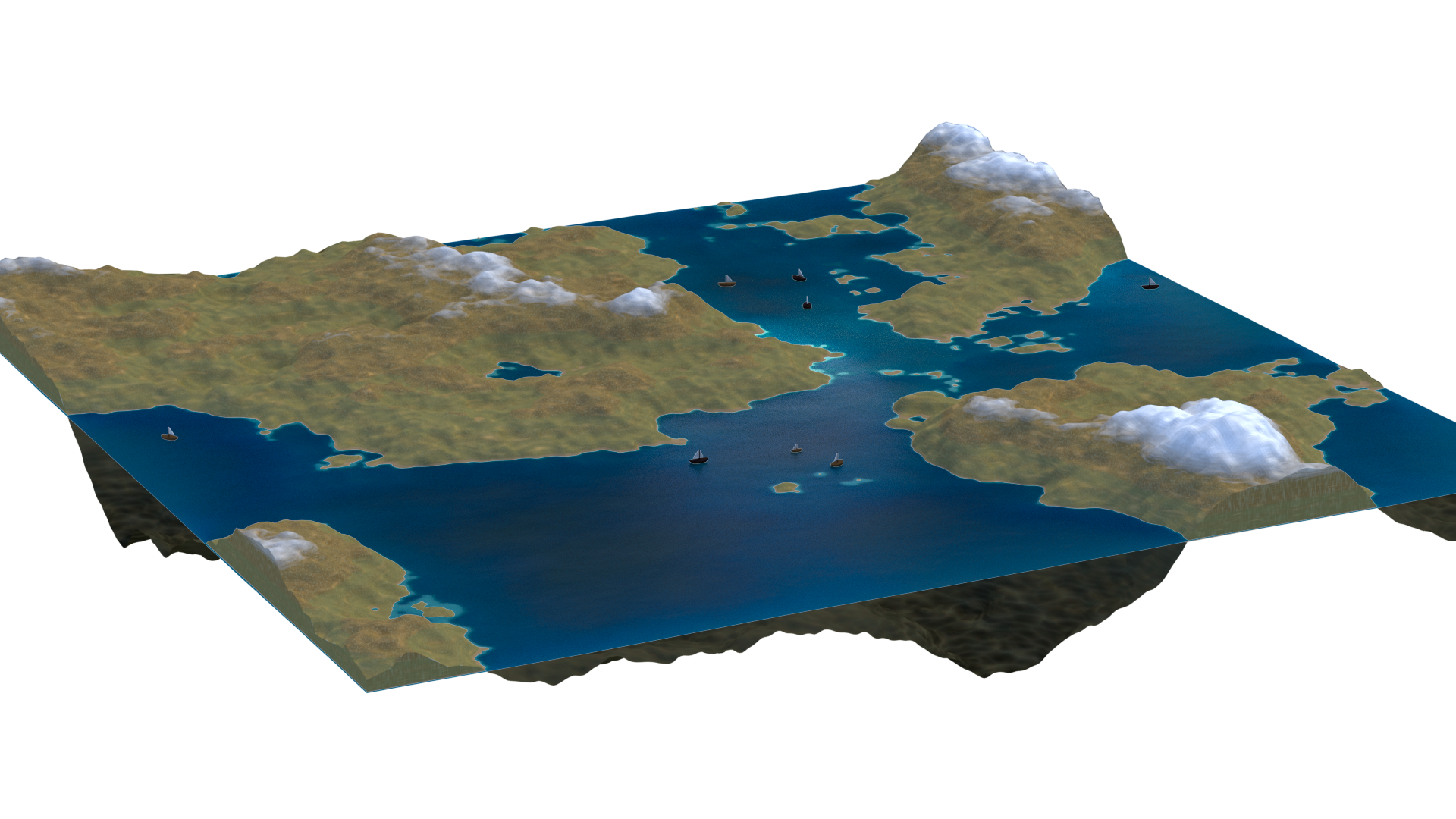}
	\caption{Raw data from zero-gradient D2M1N3 interpreted as islands.}
	\label{fig:render1}
\end{figure}

\begin{figure}[htbp]
	\centering
		\includegraphics[width=1.\textwidth]{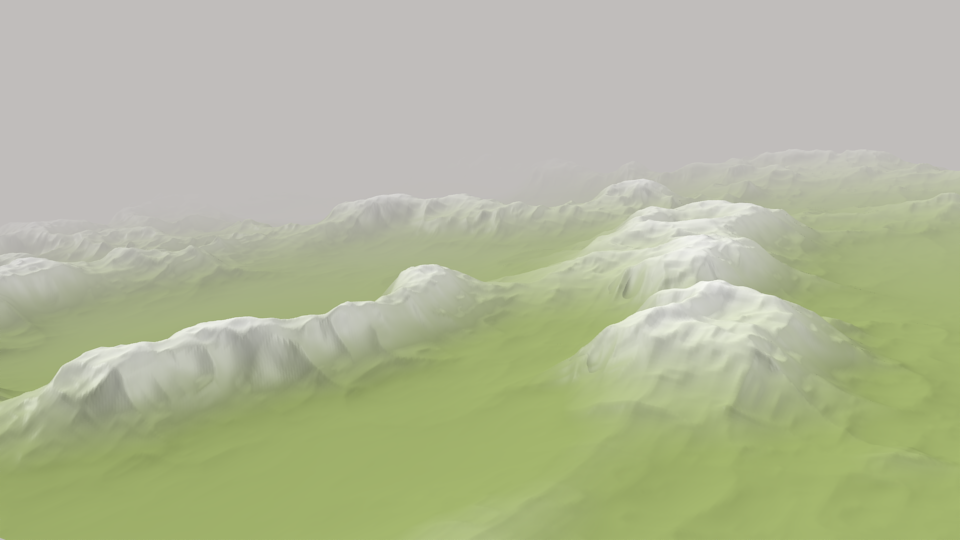}
	\caption{Ridged noise from zero-gradient D2M1N3 interpreted as mountains.}
	\label{fig:render2}
\end{figure}

\begin{figure}
    \centering
    \includegraphics[width=1.\textwidth]{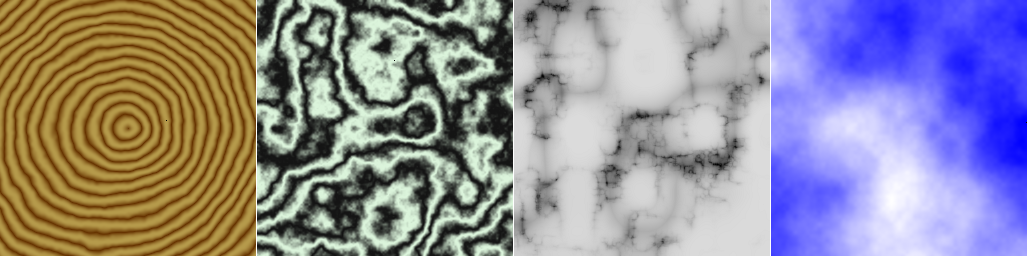}
    \caption{Examples of Wood, marble and cloud textures obtained with zero-gradient D2M1N3 model.}
		\label{fig:textures}
  \end{figure}

Figure \ref{fig:montage} depicts different ways to visualize coherent noise produced by zero-gradient D2M1N3 model, and in particular how 2D height map data is used to generate a 3D rendered mesh. Figure \ref{fig:montage} provides different types of landscapes that can be obtained with coherent noise, for each method. Again, no quality difference can visually be noted between models. Finally, Figure \ref{fig:textures} shows typical examples of coherent noise used to add turbulence to sine signal, in order to procedurally generate textures of marble, wood or stone, as first indicated in \cite{perlin85}, as well as a cloud-like texture directly obtained from the raw noise. Finally, Figure \ref{fig:render1} and Figure \ref{fig:render2} display two rendering of the terrain generated with zero-gradient D2M1N3 method, for raw noise and ridged noise \cite{libnoise} respectively.

\begin{figure}
    \centering
    \includegraphics[width=1.\textwidth]{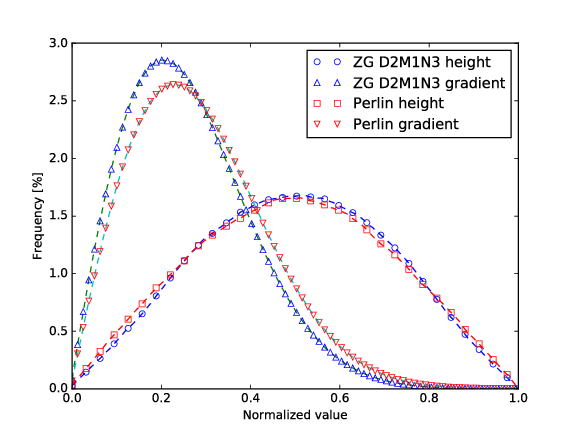}
    \caption{Comparison of the frequency distribution of the height and first gradient norm between Perlin noise and zero-gradient D2M1N3 model, for 1000 random terrains with $S=512$ pixels.}
		\label{fig:frequency}
  \end{figure}

The visual similarity between Perlin and D2M1N3 models is consistent with a zeroth and first-order gradient frequency analysis. Figure \ref{fig:frequency} shows a comparison between the two models, where the mean frequency of both height and slope norm has been obtained from 1000 random terrains with $S=512$. The height frequency distributions are very similar in both cases, while the gradient frequency distribution only slightly differ.   

%%%%%%%%%%%%%%%%%%%%% TEXTURES %%%%%%%%%%%%%%%%%%%%

%\centering
  %\begin{figure}
    %\centering
    %\includegraphics[width=.2\textwidth]{marble.png}
    %\caption{Marble texture obtained using gray colorscale and the evaluation of $sin(noise+x)$ instead of just $h$, where $x$ is a space dependant value adjusted in order to get reasonable visual result.}
  %\end{figure}
	%
  %\begin{figure}
		%\centering
    %\includegraphics[width=.2\textwidth]{wood.png}
    %\caption{Wood texture obtained using brown colorscale and the evaluation of $sin(d(x,y)+h)$ instead of just $h$, where $d(x)$ is the distance between $(x,y)$ coordinate and the center of the domain, multiplied by a constant value adjusted in order to get reasonable visual result.}
  %\end{figure}
	%\label{fig:textures}

\subsection{Performance comparison with other methods} \label{sect:results:perfs}
As seen in Section \ref{sect:noise}, D2M1N3 methods and Perlin noise have the same complexity in time, which is $\mathcal{O}(N R^2 + 2^{2N})\approx \mathcal{O}(N R^2)$ up to $N\approx 8$ in two dimensions. However, their performances may substantially differ, since constants $C_{eval}$ and $C_{init}$ are different for each method. An estimation of these values is provided here. First, examine Eqs. (\ref{c00}-\ref{c11}) first. For convenience, purely space-dependant terms are considered as having null cost, since they can be pre-computed and then accessed in constant time (this is equivalent to assume that the zoom cannot be adjusted once the execution starts). If this is not the case, they globally represent the same amount of effort in all polynomials whose order are similar, thus they can be ignored with reasonable accuracy. With these assumptions, one counts $a_g = 11$ additions and $m_g=12$ multiplications for D2M1N3 polynomial. Equation (\ref{eq:zg}) yields $a_z=3$ additions and $m_z=3$ multiplications for zero-gradient D2M1N3 polynomial. Finally, Perlin's scheme is found to include $a_p=6$ additions and $m_p=7$ multiplications. By way of comparison, examination of an efficient implementation of 2D simplex noise yields 6 additions and 10 multiplications \cite{Perlin2001}. Observing that $a_g \approx 4a_z$, $m_g \approx 4m_z$, $a_p \approx 2a_z$, $m_p \approx 2m_z$, one obtains that the ratios $C_{eval,z} / C_{eval,g} \approx 4$ and $C_{eval,z} / C_{eval,p} \approx 2$ are independant of the specific cost of addition and multiplication on the machine used. For this reason, zero-gradient D2M1N3 model is expected to run approximately twice faster than Perlin's model and four time faster than generic D2M1N3 model. This estimation tends to be weaker as the number of octaves becomes larger, as contribution of $C_{init}$ tends to be significant.

\begin{figure}
    \centering
    \includegraphics[width=1.\textwidth]{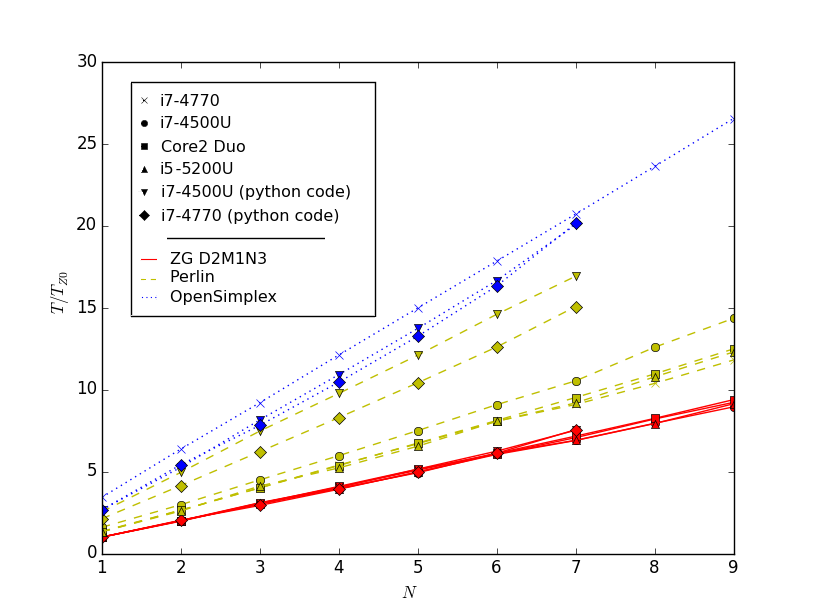}
    \caption{Measured normalized execution time as a function of the number of octaves $N$, for resolution $R=1024$ pixels. Data is presented for all methods and for all machines tested. $T$ is the measured execution time and $T_{Z0}$ is the measured execution time for zero-gradient D2M1N3 method with $N=1$ octaves. The linear behaviour show that  the computational cost is dominated by pixel evaluation and that coefficients calculation is negligible for $N<9$.}
		\label{fig:timeo}
  \end{figure}

\begin{figure}
    \centering
    \includegraphics[width=1.\textwidth]{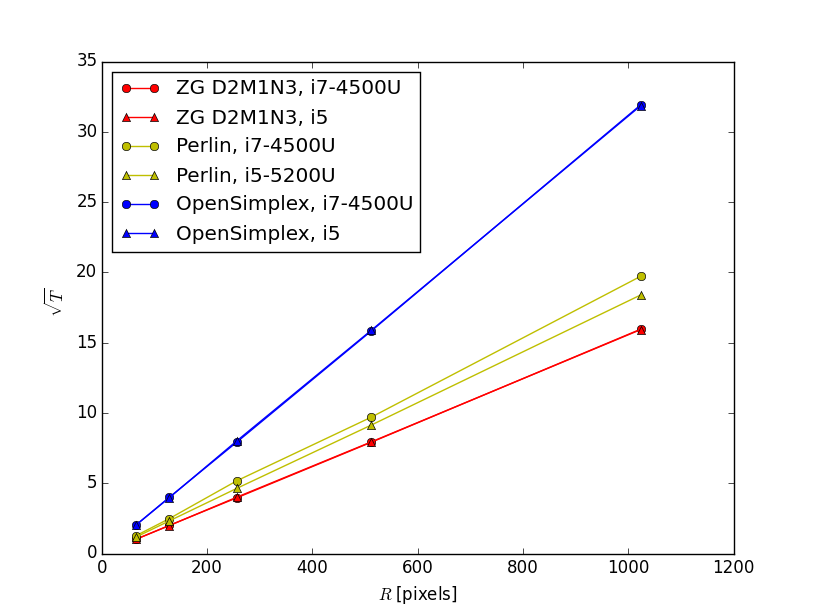}
    \caption{Example on two different machines of the square root of the measured execution time as a function of the resolution $R$, for $N=3$ octaves. The expected quadratic behaviour in $R$ shows that the computational cost is dominated by pixel evaluation $C_{eval}$ and that coefficients calculation is negligible for common resolutions.}
		\label{fig:timer}
  \end{figure}

\begin{figure}
    \centering
    \includegraphics[width=1.\textwidth]{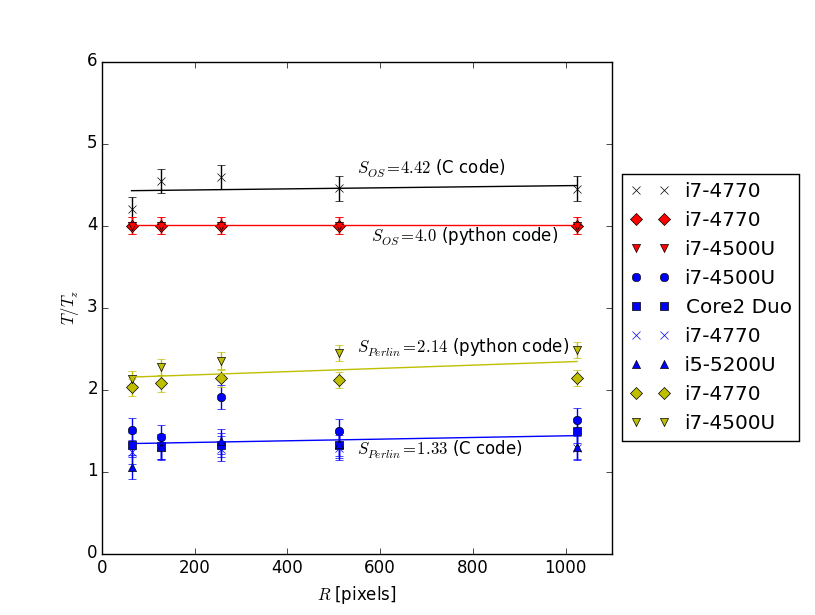}
    \caption{Speedup values of zero-gradient D2M1N3 over Perlin and OpenSimplex methods.}
		\label{fig:speedup}
  \end{figure}

Performances measurements have been done on four different home computers: Intel Core 2 Duo E6550 @ 2.33GHz (denoted Core2 Duo in the figures), Intel Core i5-5200U @ 2.20GHz, Intel Core i7-4500U @1.8GHZ, and Intel Core i7-4770 @ 3.40GHz. For each method, a C code and a Python code have been used (these codes are available as supplementary material). Execution time $T_m(N,R)$ for a given number of octaves $N$ and resolution $R$ have been obtained, for each method $m$ and with each machine, by averaging the total execution time for 1000 terrain generations. Figure \ref{fig:timeo} shows the normalized execution time for zero-gradient D2M1N3, Perlin and OpenSimplex models as a function of the number of octaves $N$. The normalized execution time is obtained by dividing the execution time $T$ by the execution time $T_{Z0}$ for $N=1$ with zero-gradient D2M1N3 method. The linear behavior of the execution time clearly indicates that the computational effort is dominated by $C_{eval}$ up to $N=9$. Figure \ref{fig:timer} displays an example of the square root of the normalized execution time, this time as a function of the resolution $R$, with a fixed number of octaves $N=3$. Again, it is evident that the cost is dominated by pixel evaluation at all tested resolutions. Finally, the time advantage of zero-gradient D2M1N3 method can be quantified by defining the speedup $S_m = T_m / T_Z$, where subscript $T_m$ is the execution time for any method $m$ and $Z$ denotes D2M1N3 method. Figure \ref{fig:speedup} shows the different speedups obtained, taking into account all the acquired data for all machines and methods. It appears that C version of zero-gradient D2M1N3 method is on average 33 \% faster than Perlin method, while Python version is more than twice faster. OpenSimplex implementations are approximately four times slower.

% Figure \ref{fig:perfratio} shows the ratio of execution time for Perlin and generic D2M1N3 models to zero-gradient model.

%   \begin{figure}
% 		\centering
%     \includegraphics[width=1.\textwidth]{perfratio_large_pix.png}
%     \caption{Ratio of the execution time for Perlin and generic D2M1N3 models to zero-gradient D2M1N3 model.}
% 		\label{fig:perfratio}
%   \end{figure}

  %\caption{Normalized performance measurements for resolution $R=1024$. Relative executiom times are in accordance with the number of operations required in the respective algorithms.}
	%\label{fig:performances}
%\end{figure}

%%%%%%%%%%%%%%%%%%%%%%%%%%%%%%%%%%%%%%%%
\subsection{Fractal analysis as a measure of realism}
The evaluation of how much a generated terrain is realistic undoubtedly depends on arbitrary choices; a person who never saw an island in his life would judge a real island as 'unrealistic' compared to all landscapes previously seen. In the other hand, it seems reasonable and intuitive to base the evaluation on quantities that can be measured both in real and numerical terrains, and which reflect in a simple fashion the complexity of a given topographical configuration. For this reason, fractal dimension is a common choice to characterize landscapes and coastlines in particular \cite{Mandelbrot,Kappraff}. It is proposed here to study the fractal dimension corresponding to the coastlines of the terrains generated with our model, and to compare it with the fractal dimension of real, natural terrains.
%This arbitrary choice is motivated by the fact that it is more convenient to find data on real coastlines than a complete heightmap of a region.

Let $L$ be the total length of a given coastline; it is clear that its value depends on the length $\epsilon$ of the measuring tape: as it becomes smaller, more details can be measured and $L$ therefore increases. This effect is known as Richardson effect and is reported in \cite{Mandelbrot}. Fractal dimension $D$ is defined as the quantity allowing to link $\epsilon$ to $L$ by a power law: $L(\epsilon) = k\cdot \epsilon^{1-D}$, with $k$ a constant. It follows that $\log(L) = (1-D) \log(\epsilon) + C$ with $C$ a constant, thus the fractal dimension can be deduced from the slope of the function $\log(L)$, whose value is $1-D$. A formal study of fractal dimension can be found in \cite{Barnsley}.

\begin{figure}
    \centering
    \includegraphics[width=1.\textwidth]{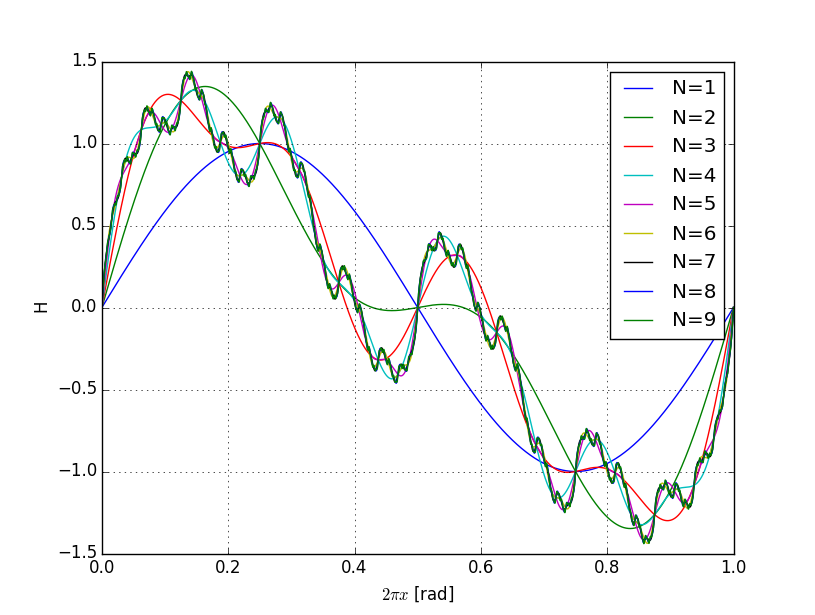}
    \caption{Superimposition $H(x)$ of sinusoids with increasing frequencies and decreasing amplitudes. Here $a = f = 2$.}
		\label{fig:fracsin}
  \end{figure}

\begin{figure}
    \centering
    \includegraphics[width=1.\textwidth]{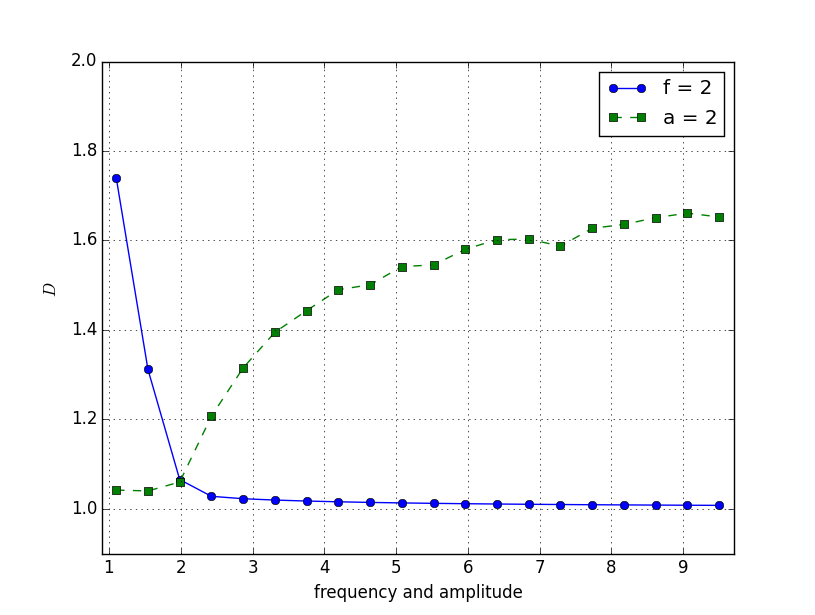}
    \caption{Fractal dimension $D$ as a function of frequency $f$ and amplitude $a$. As expected $D$ is close to 2 when $a$ is small, in other words when low octaves do not predominate much over high octaves. Reversely, $D$ is close to 1 when low octaves almost entirely determines the shape of the crest.}
		\label{fig:limfrac}
  \end{figure}

Before proceeding to a numerical measurement of $D$ for a given data set, we shall convince ourselves that a superimposition of an infinity of 1D sinusoids with exponentially decreasing amplitudes and exponentially increasing frequencies leads to a theoretical value $1\leq D \leq 2$. Sinus functions are considered in order to easily handle periodicity.The height function at octave $i$ is defined as $h_i(x) = a^{-i} \sin(f^{i}x)$, with real constants $a$ and $f$. The crest resulting from the superimposition of all $N$ octaves is
\begin{equation}
H(x) = \sum_{i=0}^N h_i(x),
\end{equation}
for $x\in [0,2\pi]$. An example of the obtained crest is shown in Figure \ref{fig:fracsin}, for different number of octaves. Using that $\partial h_i(x) / \partial x = (f/a)^i \cos(f^{i}x)$ The total length of the crest can be expressed as
\begin{equation}\label{eq:L}
L(N) = \int_0^{2\pi} \sqrt{1+\left(\sum_{i=0}^N (f/a)^i \cos(f^{i} x)\right)^2} dx,
\end{equation}
From the mathematical point of view, the actual length $L$ of the crest is the length obtained with an infinite number of octaves. In the other hand, as previously said, one seeks a relationship on the form $L(N) = k \epsilon^{1-D}$. Consequently, one obtains
\begin{equation}
D = \lim_{(N,\epsilon) \rightarrow (\infty,0)}\frac{-\log L(N)}{\log \epsilon} + 1.
\end{equation}
Because of the difficulty to solve the integral in Equation (\ref{eq:L}), a numerical evidence of the convergence value of $D$ for different choices of parameters $a$ and $f$ is provided in Figure \ref{fig:limfrac}.

\begin{figure}
\centering
\includegraphics[width=1.\textwidth]{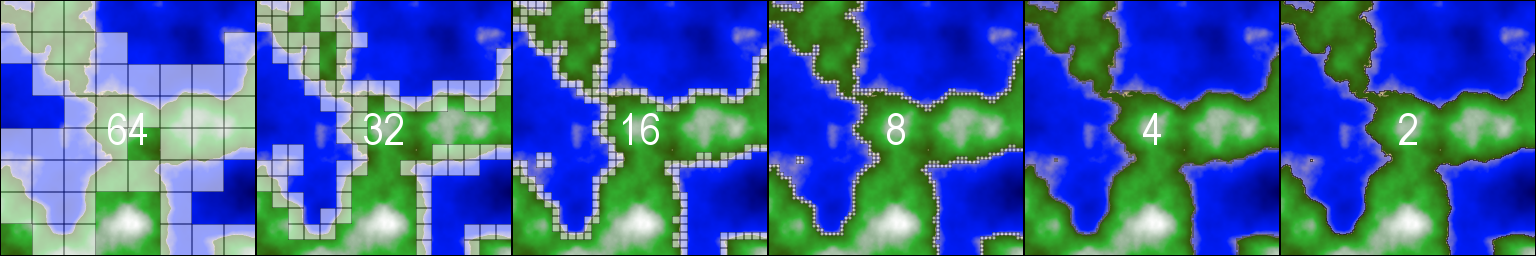}
\caption{Example of box counting process used for computing fractal dimension of a generated map. Box size varies from 2 to 64 pixel and is indicated on each image. In this example, the terrain size is $512\times 512$ pixels large.}
\label{fig:megafractal}
\end{figure}

\begin{figure}
		\centering
    \includegraphics[width=1.\textwidth]{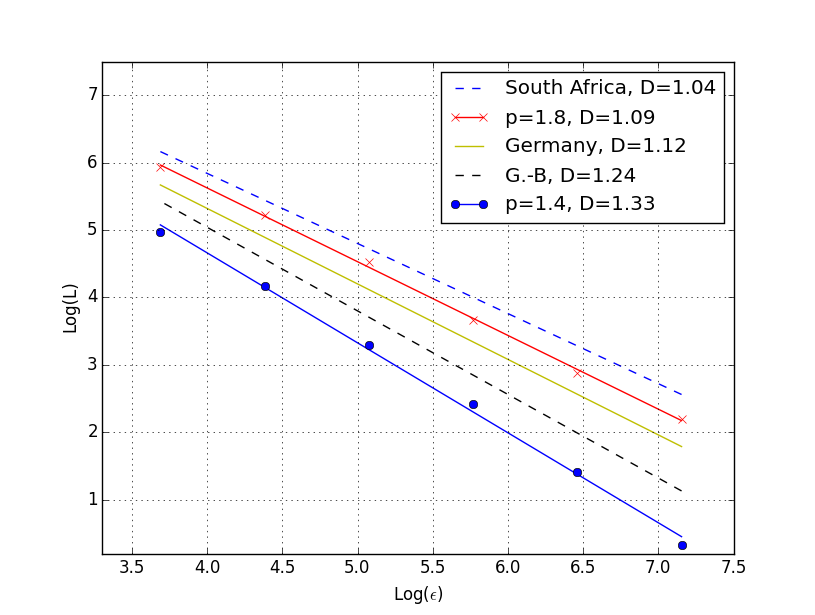}
    \caption{Example of computed fractal dimension for different values of persistence $p$ and comparison with values for South Africa, Germany and Great-Britain coastline obtained by Richardson and reported in \cite{Mandelbrot}. The plot display the logarithmic value of coastline length $L$ as a function of the logarithmic value of the box size $\epsilon$, and the fractal dimension is the corresponding gradient. Note that ordinate of the curves has been arbitrary chosen for convenience, as their slopes is the only quantity of interest here.}
		\label{fig:dexample}
  \end{figure}

\begin{figure}
    \centering
    \includegraphics[width=1.\textwidth]{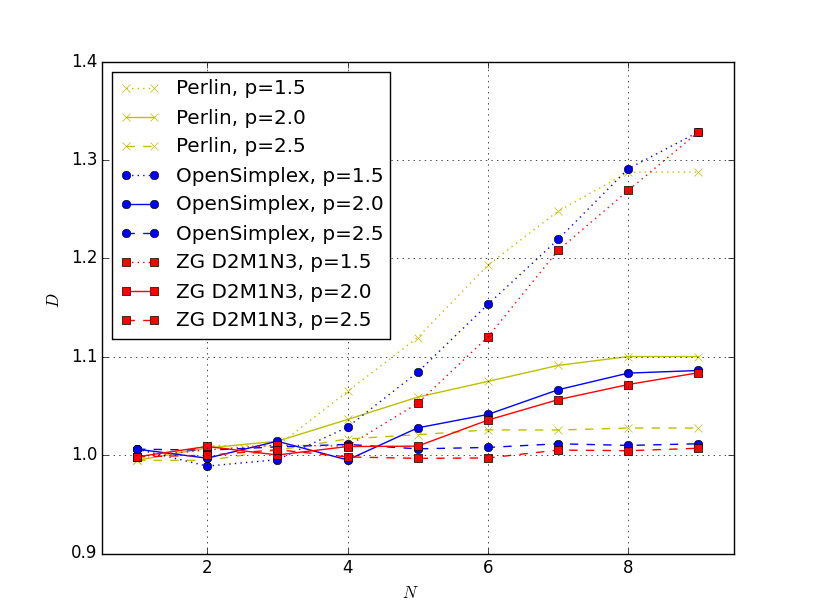}
    \caption{Computed fractal dimension $D$ as a function of the number of octaves $N$ used to generate the height map, for different values of persistence $p$. The results of zero-gradient D2M1N3 method are compared to Perlin and OpenSimplex models.}
		\label{fig:dn}
  \end{figure}

In order to measure $D$ for a given numerical landscape, each terrain is generated once and for all with a given number of octaves $N$ and a resolution $R = 1024$. Several exponentially decreasing values for $\epsilon$ are then chosen, to which are associated squares of side length $\epsilon$. By counting, for each given $\epsilon$, how many squares are needed to cover all the coastline of the height map, one is able to deduce $D$ from a linear regression of $log(L)$ plot, as explained above. Figure \ref{fig:megafractal} depicts the process, for an example map of $512 \times 512$ pixels. In Figure \ref{fig:dexample} are displayed fractal dimension reported in \cite{Mandelbrot} from Lewis Richardson's empirical work for South Africa, Germany land frontier and West coast of Britain; these experimentally found quantities are compared to data generated with zero-gradient D2M1N3 method for different values of persistence. Figure \ref{fig:dn} shows how $D$ evolves as the number of octaves used for generating the height map grows, again for different values of persistence, as expected after studying the behaviour of fractal sinusoids. One can note that the results are essentially similar between zero-gradient D2M1N3, Perlin and OpenSimplex algorithms. It should be noted that even though $D$ may continue to grow significantly, the human, visual appreciation of the result is limited at approximately 8 octaves for common model parameters, as argued below.

  %\caption{Normalized performance measurements for resolution $R=1024$. Relative executiom times are in accordance with the number of operations required in the respective algorithms.}
	%\label{fig:fractaldim}
%\end{figure}

It is worth noting that fractal dimension of coastlines is not by itself a complete measure of terrain realism; Koch snowflake has very little resemblance with real terrain, while its fractal dimension is 1.26 \cite{Kappraff}. Therefore, it seems judicious to associate $D$ with another criterium whose role is to quantify how much terrain is uneven. Although this point is not investigated, it is worth noting that the Kullback-Leibler \cite{kullback} divergence could be a suitable choice for measuring the similarity between height maps as a complement to fractal analysis.

\section{Conclusion}
Two main uses can be made of the model, either by using zero-gradient D2M1N3 model to increase performance compared to standard Perlin noise, or by using a higher-order scheme (and in particular constrained mixed derivatives) with the aim to arbitrarily constrain gradient. While the former correspond to typical video-games demand, the latter may find use in scientific and industrial research, as discussed in the introduction.

\subsection*{Strengths}
The new method described allows to significantly improve performances of realistic terrain generation based of fractal brownian motion. In addition, the general model proposed allows to reach an arbitrary level of gradient smoothness. The scheme for noise generation consists in solving, once and for all at the theoretical level, a linear set of equations whose order depends on the number of desired constraints on gradients. It has been shown that zero-gradient D2M1N3 model is the minimal configuration for smooth 3D polynomial terrain generation. Another benefit of the presented method is that, unlike simplex models, it is very similar to Perlin noise, which means that quality and performance improvements developed for Perlin noise such as \cite{ParberryTerrain,ParberryAmortized} can be straightforwardly applied to it.

\subsection*{Limitations}
The performance gain of the model is paid by the loss of intuition compared to Perlin polynomial, who can be seen as a simple interpolation of height values generated from corner gradients on a grid. Moreover, for dimensions higher than 4, simplex noise has proven to give better performances than Perlin model \cite{Perlin2001}. Finally, for generation of terrain including caves, a common approach is to use a 3D height map for which some values are interpreted as void or air; in that case, a performance study comparing zero-gradient D3M1N3 model to Perlin and simplex methods should be performed in order to determinate how the performance advantage of zero-gradient method is diminished.

% We have described a general model that allows to reach an arbitrary level of continuity, considering that the scheme for noise generation consists in solving, once and for all at the theoretical level, a linear set of equations whose order depends on the number of desired constraints. In particular, we have demonstrated that zero-gradient D2M1N3 model allows to obtain significantly better performances that traditional Perlin noise while preserving quality of the produced output, and we have shown that it is the minimal configuration for polynomial terrain generation.

% This efficiency gain is paid by the loss of intuition compared to Perlin polynomial, who can be seen as a simple interpolation of height values generated from corner gradients on a grid.

\section*{Acknowledgments}

The authors thanks Gregor Chliamovitch and Paul Albuquerque for assistance with the fractal analysis of landscapes, Mohamed Ben Belgacem for assistance with the C implementations of the models, and Lapineige for the rendering of the islands shown in Figure \ref{fig:render1}.

%% The Appendices part is started with the command \appendix;
%% appendix sections are then done as normal sections
%% \appendix

%% References with bibTeX database:

%\bibliographystyle{model3-num-names}
%\bibliography{<your-bib-database>}

\bibliographystyle{plain}
\bibliography{references}

%% Authors are advised to submit their bibtex database files. They are
%% requested to list a bibtex style file in the manuscript if they do
%% not want to use model3-num-names.bst.

%% References without bibTeX database:

% \begin{thebibliography}{00}

%% \bibitem must have the following form:
%%   \bibitem{key}...
%%

% \bibitem{}

% \end{thebibliography}

\end{document}